\documentclass[epj]{svjour}
\usepackage{latexsym}
\usepackage{amssymb}
\usepackage{amsmath}
\usepackage{amscd}
\usepackage{graphicx}
\usepackage{textcomp}
\usepackage{colortbl}
\usepackage{subcaption}
\usepackage{hyperref}
\usepackage{multirow}
\usepackage{lipsum}
\begin{document}
	
\title{The simplest parametrization of equation of state parameter in the scalar field Universe}

\author{Preeti Shrivastava\inst{1,2}, A. J. Khan\inst{1}, G. K. Goswami\inst{3}, Anil Kumar Yadav\inst{4}, J. K. Singh\inst{3} }
\institute{$^{1}$Department of Mathematics, MATS University, Raipur, CG, India\\$^{2}$Department of Mathematics, Shri Shankaracharya, Mahavidyalaya, Bhilai, CG, India\\$^{3}$Department of Mathematics, Netaji Subhas University of Technology, Delhi, India\\$^{4}$Department of Physics, United College of Engineering and Research,Greater Noida - 201310, India}




\abstract{In this paper, we have investigated a scalar field cosmological model of accelerating Universe with the simplest parametrization of equation of state parameter of the scalar field.  We used $H(z)$ data, pantheon compilation of SN Ia data and BAO data to constrained the model parameters using $\chi^{2}$ minimization technique. We obtain the present values of Hubble constant $H_{0}$ as $66.2^{+1.42}_{-1.34}$, $70.7^{+0.32}_{-0.31}$ and $67.74^{+1.24}_{-1.04}$ for $H(z)$, $H(z)$ + Pantheon and $H(z)$ + BAO respectively. Also, we have estimated the present age of the Universe in derived model $t_{0} = 14.38^{+0.63}_{-0.64}$ for joint $H(z)$ and pantheon compilation of SN Ia data which has only $0.88~\sigma$ tension with its empirical value obtained in Plank collaboration \cite{Ade/2016}. Moreover, the present values of the deceleration parameter $q_{0}$ come out to be $-0.55^{+0.031}_{-0.038}$, $-0.61^{+0.030}_{-0.021}$ and $-0.627^{+0.022}_{-0.025}$ by bounding the Universe in derived model with $H(z)$, $H(z)$ + Pantheon compilation of SN Ia and $H(z)$ + BAO data sets respectively. We also have performed the state-finder diagnostics to discover the nature of dark energy.  
	\PACS{
		{98.80.-k}  \and
		{98.80.Es}  \and
		{98.80.Cq}
		{}
	}
}

\authorrunning{Preeti Srivastava et al.}
\titlerunning{The simplest parametrization of equation of state parameter...}

\maketitle

\section{Introduction}
\label{sec:intro}
We are living in a special epoch of the cosmic history where the expansion of the Universe is not smooth or uniform but it is speeding up which leads acceleration in the current Universe. However, the exact reason of this acceleration is still unknown. In general theory of relativity, the late time acceleration of the Universe is described by inclusion of dark energy density along with matter density in Einstein's field equation \cite{Kumar/2011,Yadav/2011,Yadav/2016,Goswami/2020,Amirhashchi/2020,Goswami/2020,Kumar/2011grg,Akarsu/2010} whereas in modified theories of gravity, there are some studies which describes the current acceleration of the Universe without inclusion of dark energy component \cite{Harko/2011,Prasad/2020pramana,Yadav/2019,Sharma/2018}. The late time acceleration of the Universe has been investigated observation ally using the luminosity distance of Supernovae type Ia (SN Ia) \cite{Perlmutter/1997,Perlmutter/1998,Perlmutter/1999,Riess/2004}. In addition to SN Ia observation, other observations including baryon acoustic oscillation (BAO) \cite{Blake/2011}, the cosmic microwave background (CMB) \cite{Bennett/2003} and Plank collaboration \cite{Ade/2016} support an accelerated expansion of the Universe at present epoch. The observational estimates suggest that the pressure-less dark matter and hypothetical dark energy are two main ingredients of the Universe. However, the actual physics of these dark components of the Universe are still unknown. The simplest way to describes this acceleration of the Universe is that one has to assumed tiny cosmological constant $\Lambda$ in Einstein field equations. The pressure of $\Lambda$ is negative and equal to its energy density \cite{Weinberg/1989,Peebles/2003}. This type of cosmological model is known as $\Lambda$CDM model and it has received greatest focus for its
ability to fit most of the observational data. Despite being good consistency with observations, the $\Lambda$CDM model suffers with mainly two serious problems on theoretical ground, namely the fine-tuning and the cosmic coincidence issue. Apart from these two issues, the $\Lambda$CDM model also suffers with $H_{0}$ tension which is one of the major problem at present
time within this paradigm. $H_{0}$ tension arises due to significant standard deviation in the estimated values of $H_{0}$ from the early measurements by the Planck team \cite{Aghanim/2018} and a model independent approach \cite{Riess/2019,Riess/2020}. In Ref. \cite{Valentino/2021H0}, the authors have elaborated $H_{0}$ tension and its possible solution. \\

Another way to describe the late time acceleration of the Universe is to consider Einstein - Hilbert Lagrangian as a generic function of the Ricci scalar $R$~$(f(R))$ gravity) \cite{Capozziello/2003} or function of Ricci scalar $R$ and trace of energy momentum tensor $T$~$(f(R, T)$ gravity)\cite{Harko/2011}. In 2014, Harko \cite{Harko/2014} has studied the matter-geometry coupling of modified gravity models with thermodynamic implications. Some useful applications of $f(R, T)$ theory of gravity are given in Refs. \cite{Bhardwaj/2020,Yadav/2020pramana,Sharma/2020,Singla/2020GC,Sharma/2018,Yadav/2018IJGMMP,Yadav/2015}. Furthermore, in Refs. \cite{Hu/2007,Nojiri/2003prd}, the authors have constructed viable cosmological models in $f(R)$ theory of gravity which qualify the solar system test. Some pioneer researches in $f(R)$ gravity based on the galactic dynamic of massive test particle without inclusion of dark matter have been investigated in Refs. \cite{Capozziello/2006,Martins/2007,Boehmer/2008,Boehmer/2008jcap}. Some other modified theories of gravity such as $f(G)$ \cite{Felice/2009}, $f(R,G)$ \cite{Bamba/2010} and $f(T, B)$ \cite{Bahamonde/2018} theories have been also investigated in recent times. A wide range of phenomena can be produced from modified theories of gravity by adopting different functions. However, many functional forms are not favored by recent cosmological observations.\\

Apart from the modified theories of gravity or cosmological constant inspired models, the scalar fields with time or redshift varying equation of state are the most favored for producing acceleration in the Universe at present epoch. The scalar field acquire negative pressure during slow roll down of scalar potential $V(\phi)$ \cite{Caldwell/1998,Ferreira/1998,Wands/1998,Liddle/1999,Dodleson/2000}. The scalar field as notion of tracker potentials in quintessence theory have been introduced in Refs. \cite{Zlatev/1999,Steinhardt/1999,Johri/2002}. These tracker field induced scalar field cosmological models avoid the fine-tuning and the coincidence problems. Johari \cite{Johri/2001} has introduced the concept of integrated tracking which essentially show that the tracker potentials follow a definite path of evolution of the Universe, in compatibility with the observational constraints. Some important applications of time varying equation of state parameter are discussed in Refs. \cite{Sahni/2004LNP,Sahni/2000,Chimento/1996}. The presence of scalar field $\phi$ is also observed by several fundamental theories which motivate us to study the dynamic properties of scalar fields in cosmology. A wide range of scalar-field cosmological models has been suggested so far \cite{Hartle/1984,Hawking/1984,Vilenkin/1983,Barvinsky/1994,Spokoiny/1984,Salopek/1989,Khalatnikov/1992}. Kamenshchik et al \cite{Kamenshchik/2001} have investigated Chaplygin gas type dark energy model with peculiar equation of state parameter. \\ 

In this paper, we have considered the parametrization of equation of state parameter and obtained explicit solution of Einstein field equations in flat FRW space time. The structure of this paper is organized as: In section \ref{sec:2}, the theoretical model and its basic equations are given. In section \ref{sec:3}, we present all the details of the observational data used in this paper to constrain the cosmological parameters their uncertainties. The physical properties of the Universe in derived model are discussed in section \ref{sec:4}.  Finally, in section \ref{sec:5}, we summarize our results focusing on the main
ingredients of the model.\\
\section{Theoretical model and basic equations}
\label{sec:2}
We consider following action for Einstein's field equations in scalar field Universe.
\begin{equation}
\label{A-1}
S= S_g + S_m.
\end{equation}
where $S_{g}$ and $S_{m}$ denote action due to gravitation and matter respectively.\\
The action due to gravitation and baryon matter are respectively defined as
\begin{equation}
\label{A-2}
S_g= \int d^4x\sqrt{-g} \left[\frac{R}{16\pi G}+\left\{\frac{1}{2}g^{ij}\phi_i\phi_j-V(\phi)\right\}\right],
\end{equation}
\begin{equation}
\label{A-3}
S_m= \int \mathcal{L}\sqrt{-g} d^4x.
\end{equation}
where, $R$, $G$ and other symbols have their usual meaning.\\
Therefore, Einstein's field equation is recast as
\begin{equation}
\label{F-1}
R_{ij}-\frac{1}{2}Rg_{ij} =-8\pi G~T_{ij}-\phi_i\phi_j + g_{ij}\left(\frac{1}{2}\phi^k\phi_k-V(\phi)\right).
\end{equation}  
Also, the action $S$ is varying with respect to scalar field $\phi$ which leads following additional equation
\begin{equation}
\label{A-4}
\phi_{;i}^i+V'(\phi)=0.
\end{equation}
The energy-momentum tensor for perfect fluid distribution is read as
\begin{equation}
\label{A-5}
T_m^{ij}=(p+\rho)u^{i}u^{j}-pg^{ij}.
\end{equation}
where $g_{ij}u^{i}u^{j}=1$.\\
The FLRW space-time (in unit $c = 1$) is given by
\begin{equation}
\label{A-6}
ds{}^{2}=dt{}^{2}-a(t){}^{2}\left[dx{}^{2}+dy{}^{2}+dz{}^{2}\right].
\end{equation}
where $a(t)$ is the scale factor which defines rate of expansion along spatial direction. \\
In co-moving co-ordinates, $u^{i} = 0; ~ i = 1, 2 \;or \;3$.\\
Since, the space-time (\ref{A-6}) represents spatially homogeneous and isotropic Universe. Therefore, one can consider a time varying scalar field $i. e.$ $\phi = \phi(t)$.\\
The field Eqs. (\ref{F-1}) and (\ref{A-4}) for metric (\ref{A-6}) are read as
\begin{equation}
\label{F-2}
2\frac{\ddot{a}}{a}+H^2 = -8 \pi G\left(\frac{\dot{\phi}^2}{2}-V(\phi)\right),
\end{equation}
\begin{equation}
\label{F-3}
3H^2 = 8 \pi G \left( \rho_m + \frac{\dot{\phi}^2}{2}+V(\phi)\right),~ H = \frac{\dot{a}}{a}.
\end{equation}
\&
\begin{equation}\label{F-4}
\ddot{\phi}+3H\dot{\phi}+ V'(\phi)=0.
\end{equation} 
where $V'(\phi) = \frac{dV}{d\phi}$ and $V(\phi)$ denotes the scalar field potential.\\  
Eq.(\ref{F-4}) is recast as
\begin{equation}
\label{F-5}
\frac{d}{dt}\left[\frac{1}{2}\dot{\phi}^2+V(\phi)\right]+3\frac{\dot{a}}{a}\dot{\phi}^2=0.
\end{equation}
Thus, the energy momentum tensor of scalar field is obtained as
\begin{equation}
\label{F-6}
T_{\phi}^{ij}=(p_{\phi}+\rho_{\phi})u^{i}u^{j}-p_{\phi}g^{ij}.
\end{equation}
where $\rho_{\phi} =\frac{1}{2}\dot{\phi}^2+V(\phi)$ and $p_{\phi} =\frac{1}{2}\dot{\phi}^2-V(\phi)$.\\
Now, the equation of state parameter for scalar field is defined as $\omega_{\phi} = \frac{p_{\phi}}{\rho_{\phi}}$.\\
Hence, the scalar field potential in terms of $\omega_{\phi}$ is computed as
\begin{equation}
\label{F-7}
V(\phi)=\frac{1-\omega_{\phi}}{2(1+\omega_{\phi})}\dot{\phi}^2.
\end{equation}
From Eqs. (\ref{F-2})$ - $ (\ref{F-4}), we observe that there are systems of three equations with four $H$, $\rho_{m}$, $\phi$ and $V$ variable. Hence, one can not solve these equation in general. However, to get explicit solution of above equations, we have to assumed at least one reasonable relations among the variables or parametrize the variables. That is why, we have considered the simplest parametrization of equation of state parameter of scalar field, given by Gong and Zhang \cite{Gong/2005}
\begin{equation}
\label{P-1}
\omega_{\phi}=\dfrac{(\omega_{\phi})_0}{1+z}.
\end{equation} 
where $(\omega_{\phi})_0$ denotes the present value of equation of state parameter of scalar field. The main reason of considering parametrization of $\omega_{\phi}$ in form of Eq. (\ref{P-1}) is that, at $z = 0$, it gives $\omega_{\phi} = (\omega_{\phi})_0$ and as $z \rightarrow \infty$, $\omega_{\phi} \rightarrow 0$ which is eventually true for modeling the  Universe.\\
Using Eqs. (\ref{F-7}) and (\ref{P-1}), Eq. (\ref{F-5}) reduces to
\begin{equation}
\label{F-8}
\frac{d}{dt}\left[\frac{1}{2}\dot{\phi}^{2}+\frac{1+z-(\omega_{\phi})_0}{2[1+z+(\omega_{\phi})_0]}\right] + 3\frac{\dot{a}}{a}\dot{\phi}^{2} = 0.
\end{equation}
Integrating Eq.(\ref{F-8}), we obtain
\begin{equation}\label{8}
\dot{\phi}^2=\dot{\phi}^2_0\frac{(\omega_{\phi})_0+z+1}{(\omega_{\phi})_0+1} (z+1)^2~ exp\left[{\frac{3(\omega_{\phi})_0 z}{z+1}}\right]. 
\end{equation}
where $\dot{\phi}_{0}$ denotes the value of $\dot{\phi}$ at $z = 0$.\\
Thus, the expression for $\rho_{\phi}$ and $p_{\phi}$ are read as
\begin{equation}
\label{F-9}
\rho_{\phi} = \frac{1}{2}\dot{\phi}^2+V(\phi) = (\rho_{\phi})_{0}(z+1)^3 exp\left[^{\frac{3(\omega_{\phi})_0 z}{z+1}}\right].
\end{equation}
\begin{equation}
\label{F-10}
p_{\phi}= \omega_{\phi}\rho_{\phi} = \dfrac{(\omega_{\phi})_{0}}{1+z}\rho_{\phi}.
\end{equation}
The continuity equation is given as
\begin{equation}
\label{F-11}
\dot{\rho_{m}} + 3\rho_{m}H +\dot{\rho_{\phi}} + 3(\rho_{\phi} + p_{\phi})H = 0,
\end{equation}
Eqs. (\ref{F-5}) and (\ref{F-11}) leads to
\begin{equation}
\label{F-12}
\dot{\rho_{m}} + 3\rho_{m}H = 0.
\end{equation}
Integrating Eq. (\ref{F-12}), we obtain 
\begin{equation}\label{9}
\rho_{m}=(\rho_m)_{0}(1+z)^{3}.
\end{equation}
Here, the parameters with suffix 0 denote its present value.\\ 
From Eqs. (\ref{F-2}) and (\ref{F-3}), the expression for deceleration parameter $q$ and Hubble's parameter $H$ are respectively obtained as
\begin{equation}\label{q}
2q = 1+\frac{3(\omega_{\phi})_{0}(\Omega_{\phi})_{0}~exp\left({\frac{3(\omega_{\phi})_0 z}{z+1}}\right)}{(1+z)\left[(\Omega_{m})_{0}+(\Omega_{\phi})_{0}~exp\left({\frac{3(\omega_{\phi})_0 z}{z+1}}\right)\right]}.
\end{equation}
\begin{equation}\label{H}
H(z) = H_{0}\sqrt{(1+z)^{3}\left[(\Omega_{m})_{0}+(\Omega_{\phi})_{0} ~exp\left({\frac{3(\omega_{\phi})_0 z}{z+1}}\right)\right]}.
\end{equation}
The luminosity distance is read as
\begin{equation}
\label{L-1}
D_{L} = (1+z)\int_{0}^{z}\frac{dz}{H(z)}.
\end{equation}
Thus, the distance modulus $\mu$ is obtained as
\begin{equation}
\label{L-2}
\mu = m - M = 5log_{10}D_{L}(z) + \mu_{0}. 
\end{equation}
where $m$ and $M$ are apparent magnitude and absolute magnitude of any distant luminous object respectively. $\mu_{0} = 5log_{10}\left(H_{0}^{-1}/Mpc\right) + 25$ is the marginalized nuisance parameter.\\ 
\section{Observational constraints}\label{sec:3}  
In this section, we use $H(z)$ data sets, Pantheon compilation of SN Ia data and Baryon Acoustic Oscillation (BAO) data sets to constrain the model parameters of the Universe in derived model. Note that the complete list of $H(z)$ data points is compiled in Refs. \cite{Sharov/2018,Biswas/2019}. The pantheon compilation of SN Ia data in the redshift range $0.01 < z < 2.3$ is given in Scolnic et al. \cite{Scolnic/2018}. The third data set, we consider is the Baryon Acoustic Oscillation (BAO) data which includes six distinct measurements of the baryon acoustic scale. The BAO data points are summarized in Table I.\\
\begin{table}[ht]
\centering
Table I: The BAO data points which we use in our analysis
\setlength{\tabcolsep}{20pt}
\scalebox{0.9}{
\begin{tabular} {cccc}
\hline
S. N.  &  $z$    &  $d_{i}$   &  References\\[0.5ex]			
\hline{\smallskip}
1  &  0.106 &   0.336        & \cite{Beutler/2011}\\[0.5ex]
			
2  &  0.35 &    0.113          & \cite{Padmanabhan/2012} \\[0.5ex]
			
3  &  0.57 &    0.073       & \cite{Anderson/2013}\\[0.5ex]
			
4  &  0.44  &   0.0916          & \cite{Blake/2011a} \\[0.5ex]
			
5  &  0.60   &  0.0726           & \cite{Blake/2011a} \\[0.5ex]
			
6  &  0.73  &   0.0592           & \cite{Blake/2011b} \\[0.5ex]
\hline
\end{tabular}}
\label{tab-1}
\end{table}

To obtain, $\chi^{2}_{BAO}$, we adopt the same procedure as given in Ref. \cite{Hinshaw/2013}. Therefore, $\chi^{2}_{BAO}$ is computed as
\begin{equation}
\label{chi-1}
\chi^{2}_{BAO} = X^{T}C^{-1}_{BAO}X ,
\end{equation}
where $X = [d(0.106) - 0.336, \frac{1}{d(0.35)} - \frac{1}{0.113}, \frac{1}{d(0.57)} - \frac{1}{0.073}, d(0.44) - 0.0916, d(0.6) - 0.0726, d(0.73) - 0.0592]$,\\
and\\
$d(z) = \frac{r_{s}(z_{drag})}{D_{V}(z)}$,\\
with\\
$r_{s}(a) = \int^{a}_{0}\frac{c_{s}da}{a^{2}H(a)}$,\\
is the co-moving sound horizon at the baryon drag epoch, $c_{s}$ the baryon sound speed and $D_{V}(z)$ is defined as\\
$D_{V}(z) = \left[(1+z)^{2}D_{A}^{2}(z)\frac{z}{H(z)}\right]^{\frac{1}{3}}$,\\
here, $D_{A}(z)$ is the angular diameter distance.\\ 

The $\chi^{2}$ for $H(z)$ data is read as
\begin{equation}
\label{chi-2}
\chi^{2}_{H(z)} = \sum_{i=1}\left[\frac{H_{th}(z_{i})-H_{obs}(z_{i})}{\sigma_{i}}\right]^{2}.
\end{equation}
where $H_{th}(z_{i})$ and $H_{obs}(z_{i})$ denote the theoretical and observed values respectively, and $\sigma_{i}^{2}$ denotes standard deviation of each $H_{obs}(z_{i})$.\\
Since, these data sets are independent from one another. Therefore, the joint $\chi^{2}$ is obtained as
\begin{equation}
\label{chi-3}
\chi^{2}_{H(z) + Pantheon} = \chi^{2}_{H(z)} + \chi^{2}_{Pantheon}.
\end{equation}
and\\
\begin{equation}
\label{chi-4}
\chi^{2}_{H(z) + BAO} = \chi^{2}_{H(z)} + \chi^{2}_{BAO}.
\end{equation} 
\begin{figure}[ht]
\centering
\includegraphics[width=5.5cm,height=5.5cm,angle=0]{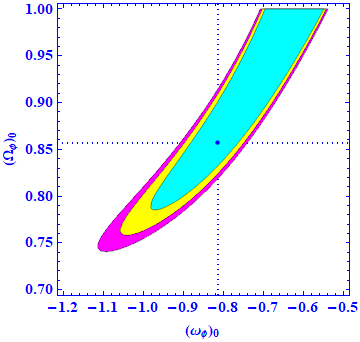}
\caption{Two dimensional contours in $(\omega_{\phi})_{0}\;-\;(\Omega_{\phi})_{0}$ plane at $1\sigma$, $2\sigma $ and $3\sigma$ confidence regions by bounding our model with $H(z)$ data.}
\end{figure}  
\begin{figure}[ht]
\centering
\includegraphics[width=5.5cm,height=5.5cm,angle=0]{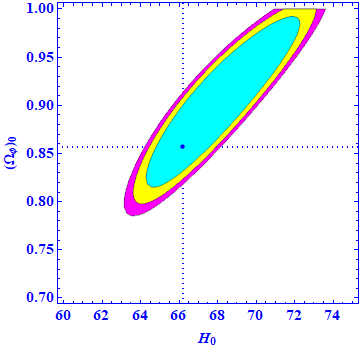}
\caption{Two dimensional contours in $H_{0}\;-\;(\Omega_{\phi})_{0}$ plane at $1\sigma$, $2\sigma $ and $3\sigma$ confidence regions by bounding our model with $H(z)$ data }
\end{figure}  
\begin{figure}[ht]
\centering
\includegraphics[width=5.5cm,height=5.5cm,angle=0]{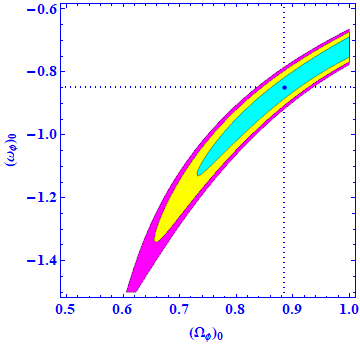}
\caption{Two dimensional contours in $(\omega_{\phi})_{0}\;-\;(\Omega_{\phi})_{0}$ plane at $1\sigma$, $2\sigma $ and $3\sigma$ confidence regions by bounding our model with joint $H(z)$ and pantheon compilation of SN Ia data.}
\end{figure}  
\begin{figure}[ht]
\centering
\includegraphics[width=5.5cm,height=5.5cm,angle=0]{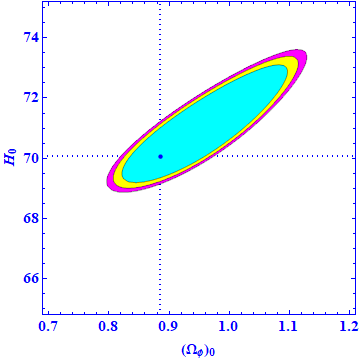}
\caption{Two dimensional contours in $(\omega_{\phi})_{0}\;-\;(\Omega_{\phi})_{0}$ plane at $1\sigma$, $2\sigma $ and $3\sigma$ confidence regions by bounding our model with joint $H(z)$ and pantheon compilation of SN Ia data.}
\end{figure}  
\begin{figure}[ht]
\centering
\includegraphics[width=5.5cm,height=5.5cm,angle=0]{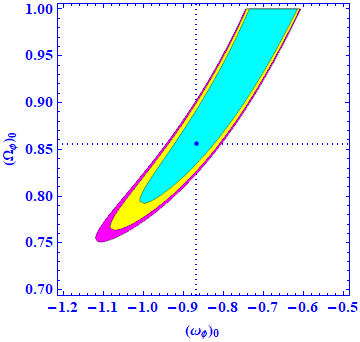}
\caption{Two dimensional contours in $(\omega_{\phi})_{0}\;-\;(\Omega_{\phi})_{0}$ plane at $1\sigma$, $2\sigma $ and $3\sigma$ confidence regions by bounding our model with joint $H(z)$ and BAO data.}
\end{figure}  
\begin{figure}[ht]
\centering
\includegraphics[width=5.5cm,height=5.5cm,angle=0]{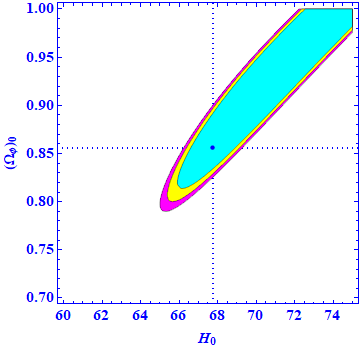}
\caption{Two dimensional contours in $(\omega_{\phi})_{0}\;-\;(\Omega_{\phi})_{0}$ plane at $1\sigma$, $2\sigma $ and $3\sigma$ confidence regions by bounding our model with joint $H(z)$ and BAO data.}
\end{figure}  
Figs. 1 - 6 depict two dimensional contours at $1\sigma$, $2\sigma $ and $3\sigma$ confidence regions by bounding our model with $H(z)$, $H(z)$ + pantheon compilation of Sn Ia data and $H(z)$ + BAO data respectively. The result of this analysis is summarized in Table II.
\begin{table}[ht]
\centering
Table I: Constrained values of model parameters
\setlength{\tabcolsep}{20pt}
\scalebox{0.7}{
\begin{tabular} {cccc}
\hline
 Parameters & $H(z)$ & $H(z)$ + Pantheon & $H(z)$ + BAO \\[1.1ex]			
\hline{\smallskip}
$H_{0}$ &  $66.2^{+1.42}_{-1.34}$ & $70.7^{+0.32}_{-0.31}$ & $67.74^{+1.24}_{-1.04}$\\[1.5ex]

$(\Omega_{\phi})_{0}$ &  $0.857^{+0.041}_{-0.025}$ & $0.856^{+0.031}_{-0.020}$ & $0.885^{+0.048}_{-0.046}$\\[1.5ex]

$(\omega_{\phi})_{0}$ &  $-0.815^{+0.066}_{-0.050}$ & $-0.869^{+0.046}_{-0.045}$ & $-0.849^{+0.028}_{-0.027}$\\[1.5ex]

\hline
\end{tabular}}
\end{table}
\section{Physical properties of the model}\label{sec:4} 
\subsection{Age of Universe}
The age of Universe in derived model is computed as
\begin{equation}
\label{age-1}
H_0 (t_0-t) = \int_{0}^{z}\frac{dz}{(1+z) h(z)};~ h(z)=H(z)/H_{0}.
\end{equation}
Therefore, the present age of the Universe is obtained as
\begin{equation}
\label{age-2}
H_0 t_0 =\lim_{z\rightarrow\infty}\int_{0}^{z}\frac{dz}{(1+z) h(z)}.
\end{equation}
where $t_{0}$ denotes the present age of the Universe.\\
\begin{figure}[ht]
\centering
\includegraphics[width=5.5cm,height=5.5cm,angle=0]{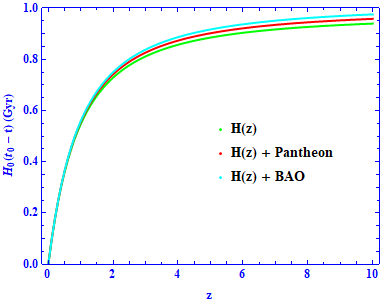}
\caption{Plot of $H_0 (t_0-t)$ versus redshift $z$.}
\end{figure}  
Fig. 7 depicts the variation of $H_{0}(t_{0} - t)$ with respect to redshift $z$. Note that we have considered the estimated values of $H_{0}$, $(\Omega_{\phi})_{0}$ and $(\omega_{\phi})_{0}$ in this paper by bounding the derived model with $H(z)$, $H(z)$ + Pantheon and $H(z)$ + BAO data sets. Integrating Eq. (\ref{age-2}) for the values of $H_{0}$, $(\Omega_{\phi})_{0}$ amd $(\omega_{\phi})_{0}$, given in table 2, we obtain present age of the Universe $t_{0}$ in this paper for $H(z)$, $H(z)$ + Pantheon compilation of SN Ia data and $H(z)$ + BAO data sets are $14.45^{+0.316}_{-0.311}$ Gyrs, $14.38^{+0.63}_{-0.64}$ Gyrs and $14.42^{+0.22}_{-0.25}$ Gyrs respectively. It is worth-whiling to note that the empirical age of the Universe extracted in Plank collaboration result \cite{Ade/2016} is given as $t_{0} = 13.81^{+0.038}_{-0.038}$ Gyrs. In some other cosmological studies, the present age of the Universe is computed as $14.46^{+0.8}_{-0.8}$ Gyrs \cite{Bond/2013}, $14.3^{+0.6}_{-0.6}$ Gyrs \cite{Masi/2002}, $14.61^{+0.22}_{-0.22}$ Gyrs \cite{Yadav/2021} and $14.5^{+1.5}_{-1.5}$ Gyrs \cite{Renzini/1996}. Thus, we observe that the age of the Universe estimated in derived model is in good agreement with its value, extracted in Plank collaboration \cite{Ade/2016}. It is important to note that the estimated age of the Universe due to joint $H(z)$ \& Pantheon compilation of SN Ia data in this paper $i. e.$ $t_{0} = 14.38^{+0.63}_{-0.64}$ has only $0.88~\sigma$ tension with Plank collaboration result \cite{Ade/2016}. Some useful remarks to the age of the Universe and its curvature are given in Ref. \cite{Valentino/2021}.  
\subsection{Deceleration parameter}
Eq. (\ref{q}) is recast as
\begin{equation}\label{q-1}
q = \frac{1}{2}\left[1+\frac{3(\omega_{\phi})_{0}(\Omega_{\phi})_{0}~exp\left({\frac{3(\omega_{\phi})_0 z}{z+1}}\right)}{(1+z)\left[(\Omega_{m})_{0}+(\Omega_{\phi})_{0}~exp\left({\frac{3(\omega_{\phi})_0 z}{z+1}}\right)\right]}\right].
\end{equation}
\begin{figure}[ht]
\centering
\includegraphics[width=2.8cm,height=2.8cm,angle=0]{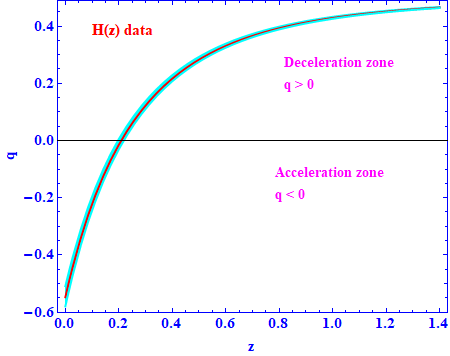}
\includegraphics[width=2.8cm,height=2.8cm,angle=0]{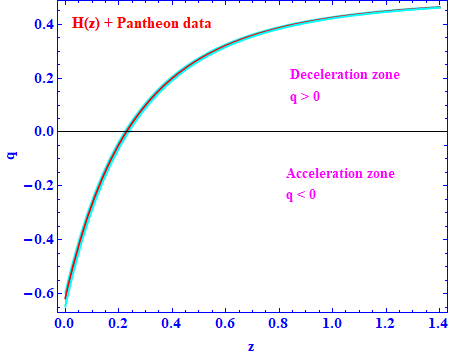}
\includegraphics[width=2.8cm,height=2.8cm,angle=0]{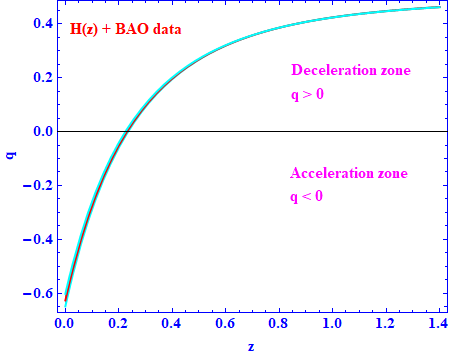}
\caption{Variation of deceleration parameter versus redshift $z$ for $H(z)$ data (left panel, $H(z)$ + Pantheon compilation of SN Ia data (middle panel) and $H(z)$ + BAO data (right panel).}
\end{figure} 
\begin{figure}[ht]
\centering
\includegraphics[width=5.5cm,height=5.5cm,angle=0]{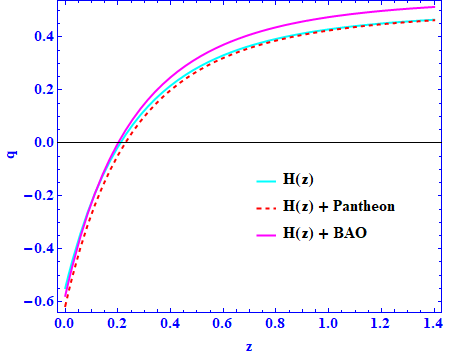}
\caption{Single plot of $q$ versus redshift $z$.}
\end{figure} 
Fig. 8 depicts the dynamics of deceleration parameter $q$ with respect to redshift $z$ for $H(z)$ data (left panel, $H(z)$ + Pantheon compilation of SN Ia data (middle panel) and $H(z)$ + BAO data (right panel). We obtain the present value of deceleration parameter $q_{0}$ as $-0.55^{+0.031}_{-0.038}$, $-0.61^{+0.030}_{-0.021}$ and $-0.627^{+0.022}_{-0.025}$ by bounding the Universe in derived model with $H(z)$, $H(z)$ + Pantheon compilation of SN Ia and $H(z)$ + BAO data sets respectively. Fig. 9 shows a single plot of $q$ versus $z$. Recently Capozziello et al \cite{Capozziello/2020} have obtained the empirical value of $q_{0}$ as $ - 0.56^{+0.04}_{-0.04}$. Some other empirical values of $q_{0}$ in the vicinity of our obtained values of $q_{0}$ are given in Ref. \cite{Cunha/2009,Jesus/2020,Singla/2021,Prasad/2021,Prasad/2021a,Prasad/2020}.   
\subsection{Statefinder diagnostics}
The statefinder pairs $\{r, s\}$ is the geometrical quantities which are directly obtained from metric. This diagnostic is used to distinguish different dark energy models and hence becomes an important tool in modern cosmology. Alam et al \cite{Alam/2003} have defined the statefinder parameters $r$ and $s$ as following
\begin{equation}
\label{S-1}
r = \frac{\dddot{a}}{aH^{3}}\;\;,\;\; s = \frac{r-1}{3(q-\frac{1}{2})}.
\end{equation} 
\begin{figure}[ht]
\centering
\includegraphics[width=5.5cm,height=5.5cm,angle=0]{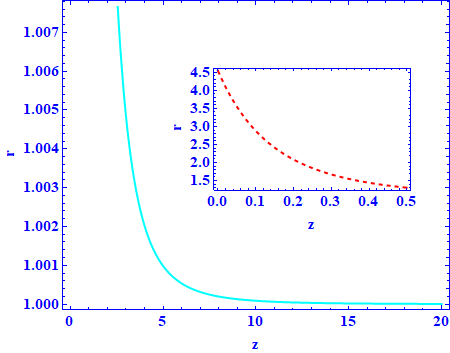}
\caption{Plot of $r$ versus $z$}
\end{figure} 
\begin{figure}[ht]
\centering
\includegraphics[width=5.5cm,height=5.5cm,angle=0]{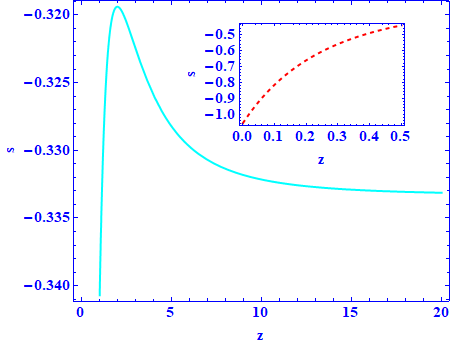}
\caption{Plot of $s$ versus $z$}
\end{figure} 
\begin{figure}[ht]
\centering
\includegraphics[width=5.5cm,height=5.5cm,angle=0]{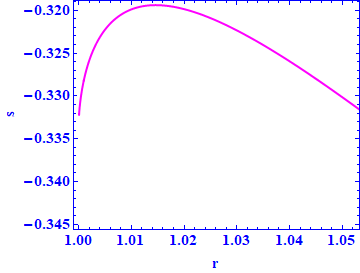}
\caption{Trajectory in $r - s$ plane}
\end{figure} 
The expression of $r$ in terms of $q$ and $z$ is obtained as
\begin{equation}
\label{S-2}
r = (2q+1)q + (1+z)\frac{dq}{dz}.
\end{equation}
Fig. 10 and Fig.11 exhibit the behaviour of $r$ and $s$ with respect to $z$ respectively. We compute $r = 0.4.54$ and $s = -1.05$ for joint $H(z)$ and pantheon compilation of SN Ia data at $z = 0$. From Figs. 10 \& 11, we observe that $r > 1$ and $s < 0$ in the redshift range $\{0, 20\}$. Also, the Universe in derived model presumes the values of statefinder pairs in the range $r > 1$ and $s < 0$ and therefore represents a Chaplygin gas type dark energy model (CGDE). We draw the temporal evolution of the Universe mimicked by our model in Fig. 12. The trajectory in $r - s$ plane clearly shows that the profile starts from the region $r > 1$ and $s < 0$ which corresponds to the CGDE Universe.   
\section{Concluding remarks}\label{sec:5} 
In this paper, we have investigated the late time accelerated expansion of the Universe by taking into account the scalar field with positive potential. To obtain an explicit solution of the field equations, we have considered the simplest parametrization of equation of state parameter $\omega_{\phi} = \frac{(\omega)_{0}}{1+z}$. This parametrization gives $\omega_{\phi} = (\omega)_{0}$ at present epoch. The scalar field potential $V(\phi)$ is directly connected to pressure through equation $p_{\phi} =\frac{1}{2}\dot{\phi}^2-V(\phi)$, therefore, the pressure $p_{\phi}$ is negative when $V(\phi) > \frac{1}{2}\dot{\phi^{2}}$ and hence $V(\phi)$ is responsible for negative pressure that leads acceleration of the Universe in the derived model. We used $H(z)$ data, pantheon compilation of SN Ia data and BAO data to constrained the model parameters using $\chi^{2}$ minimization technique. The constrained values of $H_{0}$, $(\Omega_{\phi})_{0}$ and $(\omega_{\phi})_{0}$ from all data sets are given in table 2.\\ 

Further we have also estimated the present age of the Universe as $14.45^{+0.316}_{-0.311}$ Gyrs, $14.38^{+0.63}_{-0.64}$ Gyrs and $14.42^{+0.22}_{-0.25}$ Gyrs by using $H(z)$, $H(z)$ + Pantheon compilation of SN Ia data and $H(z)$ + BAO data respectively. It is worthwhile to note that our estimated age of the Universe in derived model due to combined $H(z)$ and pantheon compilation of SN Ia data has only $0.88~\sigma$ tension in comparison with Plank collaboration results \cite{Ade/2016}. The Universe in derived model evolves with positive of deceleration parameter in its early phase of expansion and after dominance of scalar field, the Universe evolves with negative value of deceleration parameter which shows a transition from early decelerated expanding phase to current accelerated expanding phase. It is interesting to note the value of $q_{0} = -0.55^{+0.031}_{-0.038}$ obtained in our model are in good agreement with the recent results as reported in Ref. \cite{Capozziello/2020}. Furthermore, to investigate the parametrization from geometrical point of view, we also diagnose the statefinder pairs $\{r, s\}$. We observe that the Universe in derived model describes a Chaplygin gas type dark energy model (CGDE). It is worthwhile to note that the scalar field minimize $H_{0}$ tension and it is only $0.36~\sigma$ between our estimated value of $H_{0}$ along with combined $H(z)$ and BAO data and Plank collaboration result \cite{Aghanim/2018}.    


\end{document}